\documentclass[12pt]{article}
\usepackage{epsfig,latexsym,url}

\title{The Complexity of Clickomania}
\author{%
  Therese C. Biedl%
    \thanks{Department of Computer Science, University of Waterloo,
            Waterloo, Ontario N2L 3G1, Canada, email:
            \{\texttt{biedl}, \texttt{eddemaine}, \texttt{mldemaine},
              \texttt{rudolf}\}%
            \texttt{@uwaterloo.ca}.}
    \thanks{Partially supported by NSERC.}
\and
  Erik D. Demaine\footnotemark[1]
\and
  Martin L. Demaine\footnotemark[1]
\and
  Rudolf Fleischer\footnotemark[1]
\and
  Lars Jacobsen%
    \thanks{Department of Mathematics and Computer Science,
            University of Southern Denmark, Campusvej 55, DK-5230 Odense M,
            Denmark, email: \texttt{eljay@imada.sdu.dk}.
            This work was done while visiting University of Waterloo.}
\and
  J. Ian Munro\footnotemark[1] \footnotemark[2]
}
\date{}

\advance \topmargin by -\headheight
\advance \topmargin by -\headsep
\evensidemargin \oddsidemargin
\let\latexcite=\cite
\def\cite{\nolinebreak\latexcite}
\let\latexref=\ref
\def\ref{\nolinebreak\latexref}

\newtheorem{theorem}{Theorem}
\newtheorem{lemma}[theorem]{Lemma}
\newtheorem{corollary}[theorem]{Corollary}

\newenvironment{proof}{\textbf{Proof: }%
  \gdef\ProofEnding{\ifmmode ~~\Box
                    \else \hspace*{\fill}$\Box$\par \medskip
                    \fi \ignorespaces}}
  {\ProofEnding}
\newcommand\EndProofHere{\ProofEnding\gdef\ProofEnding{}}

\newcommand{\SET}[1]{\left\{#1\right\}}

\newcommand{\SIZE}[1]{\left|#1\right|}

\newcommand{\internal}{internal}

\begin{document}
\maketitle

\begin{abstract}
We study a popular puzzle game known variously as Clickomania and Same Game.
Basically, a rectangular grid of blocks is initially colored with some number
of colors, and the player repeatedly removes a chosen connected monochromatic
group of at least two square blocks, and any blocks above it fall down.
We show that one-column puzzles can be solved, i.e., the maximum
possible number of blocks can be removed, in linear time for two colors, and
in polynomial time for an arbitrary number of colors.
On the other hand, deciding whether a puzzle is solvable
(all blocks can be removed) is NP-complete for 
two columns and five colors, or five columns and three colors.
\end{abstract}

\section{Introduction}

\emph{Clickomania} is a one-player game (puzzle) with the following rules.  The
board is a rectangular grid.  Initially the board is full of square blocks each
colored one of $k$ colors.  A \emph{group} is a maximal connected monochromatic
polyomino; algorithmically, start with each block as its own group, then
repeatedly combine groups of the same color that are adjacent along an edge.
At any step, the player can select (\emph{click}) any group of size at least
two.  This causes those blocks to disappear, and any blocks stacked above them
fall straight down as far as they can (the \emph{settling} process).  Thus, in
particular, there is never an internal hole.  There is an additional twist on
the rules: if an entire column becomes empty of blocks, then this column is
``removed,'' bringing the two sides closer to each other (the \emph{column
shifting} process).

The basic goal of the game is to remove all of the blocks, or to remove as many
blocks as possible.  Formally, the basic decision question is whether a
given puzzle is \emph{solvable}: can all blocks of the puzzle be removed?
More generally, the algorithmic problem is to find the maximum number of blocks
that can be removed from a given puzzle.
We call these problems the \emph{decision} and \emph{optimization}
versions of Clickomania.

There are several parameters that influence the complexity of
Clickomania.  One obvious parameter is the number of colors.  For
example, the problem is trivial if there is only one color, or every
block is a different color.  It is natural to ask whether there is
some visible difference, in terms of complexity, between a constant
number of colors and an arbitary number of colors, or between one
constant number of colors and another.  We give a partial answer by
proving that even for just three colors, the problem is NP-complete.
The complexity for two colors remains open.

Other parameters to vary are the number of rows and the number of
columns in the rectangular grid.  A natural question is whether enforcing
one of these dimensions to be constant changes the complexity of the
problem.  We show that even for just two columns, the problem is
NP-complete, whereas for one column (or equivalently, one row), the
problem is solvable in polynomial time.  It remains open precisely how
the number of rows affects the complexity.

\subsection{History}

The origins of Clickomania seem unknown.
We were introduced to the game by Bernie Cosell~\cite{Cosell-2000-clickomania},
who suggested analyzing the strategy involved in the game.
In a followup email, Henry Baker suggested the idea of looking at a
small constant number of colors.
In another followup email, Michael Kleber pointed out that the game is
also known under the title ``Same Game.''

Clickomania! is implemented by Matthias Schuessler in a freeware
program for Windows, available from
\url{http://www.clickomania.ch/click/}.
On the same web page, you can find versions for the
Macintosh, Java, and the Palm Pilot.
There is even a ``solver'' for the Windows version, which appears to be
based on a constant-depth lookahead heuristic.

\subsection{Outline}

The rest of this paper is outlined as follows.  Section \ref{One Column in
Polynomial Time} describes several polynomial-time algorithms for the
one-column case.  Section \ref{Hardness for 5 Colors and 2 Columns} proves that
the decision version of Clickomania is NP-complete for $5$ colors and $2$
columns.  Section \ref{Hardness for 3 Colors and 5 Columns} gives the much more
difficult NP-completeness proof for $2$ colors and $5$ columns.
We conclude in Section \ref{Conclusion} with a discussion of two-player
variations and other open problems.

\section{One Column in Polynomial Time}
\label{One Column in Polynomial Time}

In this section we describe polynomial-time algorithms for the decision version
and optimization version of one-column Clickomania (or equivalently, one-row
Clickomania).  In this context, a group with more than $2$ blocks is equivalent
to a group with just $2$ blocks, so in time linear in the number of blocks
we can reduce the problem to have size linear in the number of groups, $n$.

First, in Section \ref{Reducing Optimization to Decision}, we show how to
reduce the optimization version to the decision version by adding a factor of
$O(n^2)$.  Second, in Section \ref{General One Column}, we give a general
algorithm for the decision question running in $O(k n^3)$ where $k$ is the
number of colors, based on a context-free-grammer formulation.  Finally, in
Section \ref{One-Column Two-Color Solvability in Linear Time}, we improve this
result to $O(n)$ time for $k=2$ colors, using a combinatorial characterization
of solvable puzzles for this case.

\subsection{Reducing Optimization to Decision}
\label{Reducing Optimization to Decision}

If a puzzle is solvable, the optimization version is equivalent to the decision
version (assuming that the algorithm for the decision version exhibits a valid
solution, which our algorithms do).  If a puzzle is not solvable, then there
are some groups that are never removed.  If we knew one of the groups that is
not removed, we would split the problem into two subproblems, which would be
independent subpuzzles of the original puzzle.

Thus, we can apply a dynamic-programming approach.  Each subprogram is a
consecutive subpuzzle of the puzzle.  We start with the solvable cases, found
by the decision algorithm.  We then build up a solution to a larger puzzle by
choosing an arbitrary group not to remove, adding up the scores of the two
resulting subproblems, and maximizing over all choices for the group not to
remove.  If the decision version can be solved in $d(n,k)$ time, then this
solution to the optimization version runs in $O(n^2 d(n,k) + n^3)$ time.
It is easy to see that $d(n,k) = \Omega(n)$, thus proving

\begin{lemma} \label{reduction lemma}
If the decision version of one-column Clickomania can be solved in $d(n,k)$
time, then the optimization version can be solved in $O(n^2 d(n,k))$ time.
\end{lemma}

\subsection{A General One-Column Solver}
\label{General One Column}

In this section we show that one-column Clickomania
reduces to parsing context-free languages.
Because strings are normally written left-to-right and not
top-down, we speak about one-row Clickomania in this subsection,
which is equivalent to one-column Clickomania.
We can write a one-row $k$-color
Clickomania puzzle as a word over the alphabet $\Sigma = \{c_1, \dots,
c_k\}$.
Such words and Clickomania puzzles are in one-to-one correspondence,
so we use them interchangably.

Now consider the following context-free grammar $G$:
  $$ G: ~~~ S \;\rightarrow 
  \begin{array}[t]{l}
    \Lambda \;|\; SS\; | \\
    c_iSc_i\; |\; c_iSc_iSc_i \quad \forall\; i\in\SET{1,2,\dots,k}
  \end{array}$$
We claim that a word can be parsed by this grammar precisely if it is
solvable.

\begin{theorem} \label{CFG}
The context-free language $L(G)$ is exactly the language of solvable
one-row Clickomania puzzles.
\end{theorem}

Any solution to a Clickomania puzzle
can be described by a sequence of moves (clicks), $m_1, m_2, \dots m_s$,
such that after removing $m_s$ no blocks remain.
We call a solution \emph{\internal} if the leftmost and rightmost
blocks are removed in the last two moves (or the last move, if they
have the same color).
Note that in an \internal\ solution we can choose whether to remove
the leftmost or the rightmost block in the last move.

\begin{lemma}\label{lem:left}
Every solvable one-row Clickomania puzzle has an \internal\ solution.
\end{lemma}

\begin{proof}
  Let $m_1, \dots, m_{b-1},m_b,m_{b+1}, \dots, m_s$ be a solution to a
  one-row Clickomania puzzle, and suppose that the leftmost block is
  removed in move $m_b$.  Because move $m_b$ removes the leftmost group, it
  cannot form new clickable groups.  The sequence $m_1, \dots, m_{b-1},
  \allowbreak m_{b+1}, \allowbreak \dots, m_s$
  is then a solution to the same puzzle except perhaps for the
  group containing the leftmost block. If the leftmost block is
  removed in this subsequence, continue discarding moves from the sequence
  until the remaining subsequence removes all but the group containing the
  leftmost block.  Now the puzzle can be solved by adding one more move,
  which removes the last group containing the leftmost block.
  Applying the same argument to the rightmost block proves the lemma.
\end{proof}

We prove Theorem \ref{CFG} in two parts:

\begin{lemma} \label{dir1}
If $w \in L(G)$, then $w$ is solvable.
\end{lemma}

\begin{proof}
  Because $w \in L(G)$, there is a derivation $S \Rightarrow^* w$.  The proof
  is by induction on the length $n$ of this derivation.  In the base case,
  $n=1$, we have $w=\Lambda$, which is clearly solvable.  Assume all strings
  derived in at most $n-1$ steps are solvable, for some $n\ge2$.
  Now consider the first step in a $n$-step derivation.
  Because $n \geq 2$, the first production
  cannot be $S\;\rightarrow\;\Lambda$.
  So there are three cases.

  \begin{itemize}
  \item
  $S\Rightarrow{}SS\Rightarrow^*w$:

  In this case $w = xy$, such that
  $S\Rightarrow^*x$ and $S\Rightarrow^*y$ both in at most $n-1$
  steps. By the induction hypothesis, $x$ and $y$ are solvable.
  By Lemma~\ref{lem:left}, there are \internal\ solutions for $x$ and
  $y$, where the rightmost block of $x$ and the leftmost block of $y$
  are removed last, respectively.
  Doing these two moves at the very end, we can now arbitrarily
  merge the two move sequences for $x$ and $y$, removing all
  blocks of $w$.

  \item
  $S\Rightarrow{}c_iSc_i\Rightarrow^*w$:

  In this case $w=c_ixc_i$,
  such that $S\Rightarrow^*x$ in at most $n-1$ steps. By the induction
  hypothesis, $x$ is solvable.
  By Lemma~\ref{lem:left}, there is an \internal\ solutions for $x$;
  if either the leftmost or rightmost block of $x$ has color $i$,
  it can be chosen to be removed in the last move.
  Therefore, the solution for $x$ followed by removing the
  remaining $c_ic_i$ (if it still exists) is a solution to $w$.

  \item
  $S\Rightarrow{}c_iSc_iSc_i\Rightarrow^*w$:

  This case is analogous to the previous case.
  \EndProofHere
  \end{itemize}
\end{proof}

\begin{lemma} \label{dir2}
If $w \in \Sigma^*$ is solvable, then $w \in L(G)$.
\end{lemma}

\begin{proof}
  Suppose $w\in\Sigma^*$ be solvable. We will prove that
  $w\in{}L(G)$ by induction on $\SIZE{w}$. The base case, $\SIZE{w}=0$
  follows since $\Lambda\in{}L(G)$. Assume all solvable strings of
  length at most $n-1$ are in $L(G)$, for some $n\ge1$.
  Consider the case $\SIZE{w}=n$.

  Since $w$ is solvable, there is a first move in a solution to $w$,
  let's say removing a group $c_i^m$ for $m\geq2$.
  Thus, $w=xc_i^my$.
  Now, neither the last symbol of
  $x$ nor the first symbol of $y$ can be $c_i$.
  Let $w^\prime=xy$.
  Since $\SIZE{w^\prime}\leq\SIZE{w}-2 = n-2$,
  and $w^\prime$ is solvable,
  $w^\prime$ is in $L(G)$ by the induction hypothesis.

  Observe that $c_i^m\in{}L(G)$ by one of the
  derivations:
  $$S\ \Rightarrow^{\frac{m-3}{2}}\ c_i^{\frac{m-3}{2}}Sc_i^{\frac{m-3}{2}}
  \ \Rightarrow{}\ c_i^{\frac{m-3}{2}}Sc_iSc_i^{\frac{m-3}{2}}
  \ \Rightarrow^2\ c_i^m$$
  if $m$ is odd, or
  $$S\ \Rightarrow^{\frac{m}{2}}
  \ c_i^{\frac{m}{2}}Sc_i^{\frac{m}{2}}\ \Rightarrow{}\ c_i^m$$
  if $m$ is even.
  Thus, if $x=\Lambda$, $w$ can be derived as $S\Rightarrow
  SS\Rightarrow^* c_i^mS\Rightarrow^*c_i^my=w$. Analogously for
  $y=\Lambda$.
  It remains to consider the case $x,y\neq\Lambda$.

  Consider the first step in a derivation for $w^\prime$.
  There are three cases.
  
  \begin{itemize}
  \item
  $S\Rightarrow SS\Rightarrow^* uS \Rightarrow^* uv=w^\prime$:

  We can assume that $u,v\neq\Lambda$,
  otherwise we consider the derivation of $w^\prime$
  in which this first step is skipped.
  By Lemma~\ref{dir1}, $u$ and $v$ are both
  solvable. Consider the substring $c_i^m$ of $w$ that was removed in
  the first move.
  Either $w=u_1c_i^mu_2v$ ($u_2$ possibly empty) or
  $w=uv_1c_i^mv_2$ ($v_1$ possibly empty).
  Without loss of generality, we assume the former case,
  i.e., $u=u_1u_2$.
  Then $u^\prime=u_1c_i^mu_2$ is solvable because $u$ is solvable
  and $m$ was maximal.
  Since $v\neq\Lambda$, it follows that
  $\SIZE{u^\prime}<\SIZE{w}$, and by the induction hypothesis,
  $u^\prime\in{}L(G)$.
  Hence $S\Rightarrow{}SS\Rightarrow^* u^\prime S\Rightarrow^*
  u^\prime v=w$ is a
  derivation of $w$ and $w\in{}L(G)$.

  \item
  $S\Rightarrow{}c_jSc_j\Rightarrow^*c_juc_j=w^\prime$:

  Since
  $x,y\not=\Lambda$, it must be the case that $w=c_ju_1c_i^mu_2c_j$,
  where $u=u_1u_2$.
  By Lemma~\ref{dir1}, $u$ is solvable, hence so is
  $u^\prime=u_1c_i^mu_2$ because $m$ was maximal.
  Moreover,
  $\SIZE{u^\prime}=\SIZE{w}-2$ and thus $u^\prime\in{}L(G)$
  by the induction hypothesis and
  $S\Rightarrow{}c_jSc_j\Rightarrow^*c_ju^\prime c_j=w\in{}L(G)$.

  \item
  $S\Rightarrow{}c_jSc_jSc_j\Rightarrow^*c_juc_jvc_j=w^\prime$:

  Since $x,y\not=\Lambda$,
  either $w=c_ju_1c_i^mu_2c_jvc_j$ and $u=u_1u_2$, or
  $w=c_juc_jv_1c_i^mv_2c_j$ and $u=v_1v_2$.
  Without loss of generality, assume $w=c_ju_1c_i^mu_2c_jvc_j$.
  Analogously to the previous case,
  $u^\prime=u_1c_i^mu_2\in{}L(G)$, hence
  $S\Rightarrow{}c_jSc_jSc_j\Rightarrow^*c_ju^\prime c_jvc_j=w\in{}L(G)$.
  \EndProofHere
  \end{itemize}
\end{proof}
  
Thus, deciding if a one-row Clickomania puzzle
is solvable reduces to deciding if the string $w$
corresponding to the Clickomania puzzle is in $L(G)$.
Since deciding $w\in{}L(G)$ is in $P$, so is deciding if a
one-row Clickomania is solvable.
This completes the proof of Theorem~\ref{CFG}.
In particular, we can obtain a polynomial-time algorithm for
one-row Clickomania by applying standard parsing algorithms
for context-free grammars.

\begin{corollary}
We can decide in $O(k n^3)$ time whether
a one-row (or one-column) $k$-color Clickomania puzzle is solvable.
\end{corollary}

\begin{proof}
The context-free grammar can be converted into a grammar in Chomsky normal form
of size $O(k)$ and with $O(1)$ nonterminals.  The algorithm in
\cite[Theorem~7.14, pp.~240--241]{Sipser-1997} runs in time $O(n^3)$ times
the number of nonterminals plus the number of productions,
which is $O(k)$.
\end{proof}

Applying Lemma \ref{reduction lemma}, we obtain

\begin{corollary}
One-row (or one-column)
$k$-color Clickomania can be solved in $O(k n^5)$ time.
\end{corollary}

%%%%%%%%%%%%%%%%%%%%%%%%%%%%%%%%%%%%%%%%%%%%%%%%%%%%%%%%%%%%%%%%%%%%%%%%
%\subsection{One-Column Two-Color Solvability in Linear Time}
\subsection{A Linear-Time Algorithm for Two Colors}
\label{One-Column Two-Color Solvability in Linear Time}
%%%%%%%%%%%%%%%%%%%%%%%%%%%%%%%%%%%%%%%%%%%%%%%%%%%%%%%%%%%%%%%%%%%%%%%%

In this section, we show how to decide solvability of a one-column two-color
Clickomania puzzle in linear time.  To do so, we give necessary and sufficient
combinatorial conditions for a puzzle to be solvable.  As it turns out, these
conditions are very different depending on whether the number of groups in the
puzzle is even or odd, with the odd case being the easier one.

We assume throughout the section that the groups are named $g_1,\dots,g_n$.
A group with just one block is called a \emph{singleton},
and a group with at least two blocks in it is called a
\emph{nonsingleton}.

The characterization is based on the following simple notion.  A
\emph{checkerboard} is a maximal-length sequence of consecutive groups each of
size one.  For a checkerboard $C$, $|C|$ denotes the number of singletons
it contains.
The following lemma formalizes the intuition that if a puzzle has a
checkerboard longer than around half the total number of groups, then the
puzzle is unsolvable.

\begin{lemma} \label{necessity}
Consider a solvable one-column two-color Clickomania puzzle with $n$ groups,
and let $C$ be the longest checkerboard in this puzzle.
  \begin{enumerate}
  \item If $C$ is at an end of the puzzle, then
        $|C| \leq {n-1 \over 2}$.
  \item If $C$ is strictly interior to the puzzle, then
        $|C| \leq {n-2 \over 2}$.
  \end{enumerate}
\end{lemma}

\begin{proof}
\begin{enumerate}
\item
  Each group $g$ of the checkerboard $C$ must be removed.
  This is only possible if $g$ is merged with some other group of the
  same color not in $C$, so there are at least $|C|$ groups
  outside of $C$.
  These groups must be separated from $C$ by at least one extra group.
  Therefore, $n \geq 2|C|+1$ or $|C| \leq {n-1\over 2}$.

\item
  Analogously,
  if $C$ is not at one end of the puzzle,
  then there are two extra groups at either end of $C$.
  Therefore, $n \geq 2|C|+2$ or $|C| \leq {n-2\over 2}$.
  \EndProofHere
\end{enumerate}
\end{proof}

\subsubsection{An Odd Number of Groups}

The condition in Lemma \ref{necessity} is also sufficient if the number of
groups is odd (but not if the number of groups is even).  The idea is to focus
on the \emph{median} group, which has index $m={n+1 \over 2}$.  This is
motivated by the following fact:

\begin{lemma} \label{median}
If the median group has size at least two, then the puzzle is solvable.
\end{lemma}

\begin{proof}
Clicking on the median group removes that piece and merges its two
neighbors into the new median group (it has two neighbors because $n$ is odd).
Therefore, the resulting puzzle again has a median group with size at least
two, and the process repeats.  In the end, we solve the puzzle.
\end{proof}

\begin{theorem}
A one-column two-color Clickomania puzzle with an odd number of groups, $n$,
is solvable if and only if
\begin{itemize}
\item the length of the longest checkerboard is at most $(n-3)/2$; or
\item the length of the longest checkerboard is exactly $(n-1)/2$, and
      the checkerboard occurs at an end of the puzzle.
\end{itemize}
\end{theorem}

\begin{proof}
If the puzzle contains a checkerboard of length at least $m = {n+1 \over 2}$,
then it is unsolvable by Lemma~\ref{necessity}.
If the median has size at least two, then we are also done by
Lemma~\ref{median}, so we may assume that the median is a singleton.
Thus there must be a nonsingleton somewhere
to the left of the median that is not the leftmost group,
and there must be a nonsingleton to the
right of the median that is not the rightmost group.
Also, there are two such nonsingletons with at most $n-2\over 2$
other groups between them.

Clicking on any one of these nonsingletons destroys two groups
(the clicked-on group disappears, and its two neighbors merge).
The new median moved one group right [left] of the old one if we clicked
on the nonsingleton left [right] of the median.
The two neighbors of the clicked nonsingleton merge into
a new nonsingleton, and this new nonsingleton is one closer to the other
nonsingleton than before.
Therefore, we can continue applying this procedure until the median
becomes a nonsingleton and then apply Lemma~\ref{median}.
Note that if one of the two nonsingletons ever reaches the end of the
sequence then the other singleton must be the median.
\end{proof}

Note that there is a linear-time algorithm implicit in the proof
of the previous lemma, so we obtain the following corollary.

\begin{corollary}
\label{twocolor_odd}
One-column two-color Clickomania with $n$ groups
can be decided in time $O(n)$ if $n$ is odd.
If the problem is solvable,
a solution can also be found in time $O(n)$. 
\end{corollary}

\subsubsection{An Even Number of Groups}

The characterization in the even case reduces to the odd case, by showing that
a solvable even puzzle can be split into two solvable odd puzzles.

\begin{theorem}
A one-column two-color Clickomania puzzle, $g_1, \dots, g_n$, with $n$
even is solvable if and only if there is an odd
index $i$ such that $g_1,\dots,g_i$ and $g_{i+1},\dots,g_n$ are
solvable puzzles.% (each with an odd number of groups).
\end{theorem}

\begin{proof}
Sufficiency is 
a straightforward application of Lemma \ref{lem:left}.
First solve the instance $g_1,\dots,g_i$
so that all groups but $g_i$ disappear and $g_i$
becomes a nonsingleton.
Then solve instance $g_{i+1},\dots,g_n$ so that
all groups but $g_{i+1}$ disappear
and $g_{i+1}$ becomes a nonsingleton.
These two solutions can be executed independently
because $g_i$ and $g_{i+1}$ form a ``barrier.''
Then $g_i$ and $g_{i+1}$ can be clicked to solve the puzzle.

For necessity, assume that $m_1,\dots,m_l$ is a sequence of clicks that solves
the instance.  One of these clicks, say $m_j$, removes the blocks of group
$g_1$.  (Note that this group might well have been merged with other groups
before, but we are interested in the click that actually removes the blocks.)
Let $i$ be maximal such that the blocks of group $g_i$ are also removed during
click $m_j$.

Clearly $i$ is odd, since groups $g_1$ and $g_i$ have the same color and
we have only two colors.  It remains to show that the instances
$g_1,\dots,g_i$ and $g_{i+1},\dots,g_n$ are solvable.

The clicks $m_1,\dots,m_{j-1}$ can be distinguished into two kinds: those
that affect blocks to the left of $g_i$, and those that affect blocks to
the right of $g_i$.  (Since $g_i$ is not removed before $m_j$, 
a click cannot be of both kinds.)

Consider those clicks that affect blocks to the left of $g_i$, and apply
the exact same sequence of clicks to instance $g_1,\dots,g_i$.  Since $m_j$
removes $g_1$ and $g_i$ at once, these clicks must have removed all
blocks $g_2,\dots,g_{i-1}$.  They also merged $g_1$ and $g_i$, so that
this group becomes a nonsingleton.  One last click onto $g_i$ hence gives
a solution to instance $g_1,\dots,g_i$.

Consider those clicks before $m_j$ that affect blocks to the right of $g_i$.
None of these clicks can merge $g_i$ with a block $g_k$, $k> i$, since this 
would contradict the definition of $i$.  Hence it does not matter whether
we execute these clicks before or after $m_j$, as they have no effect on
$g_i$ or the blocks to the left of it.  

If we took these clicks to the right of $g_i$, and combine them with the
clicks after $m_j$ (note that at this time, block $g_i$ and everything
to the left of it is gone), we obtain a solution to the instance
$g_{i+1},\dots,g_n$.  This proves the theorem.
\end{proof}

Using this theorem, it is possible to decide in linear time whether an
even instance of one-column two-color Clickomania is solvable, though the
algorithm is not as straightforward as in the odd case.  The
idea is to proceed in two scans of the input.  In the first scan,
in forward order, we determine for each odd index $i$ whether $g_1,\dots,
g_i$ is solvable.  We will explain below how to do this in amortized
constant time.   In the second scan, in backward order, we determine for 
each odd index $i$ whether $g_{i+1},\dots,g_n$ is solvable.  If any
index appears in both scans, then we have a solution, otherwise there
is none.

So all that remains to show is how to determine whether $g_1,\dots,g_i$
is solvable in amortized constant time.  (The procedure is similar
for the reverse scan.)  Assume that we are considering group $g_i$,
$i=1,\dots,n$.  Throughout the scan we maintain three indices, $j$, $k$ 
and $l$.  We use $j$ and $k$ to denote the current longest
checkerboard from $g_j$ to $g_k$.    Index $l$ is the minimal index such
that $g_l,\dots,g_{i}$ is a checkerboard.  We initialize $i=j=k=l=0$.

When considering group $g_i$, we first update $l$.  If $g_i$ is a singleton,
then $l$ is unchanged.  Otherwise, $l=i+1$.  Next, we update $j$ and $k$,
by verifying whether $i-l>k-j$, and if so, setting $j=l$ and $k=i$.
Clearly, this takes constant time.

For odd $i$, we now need to verify whether the instance $g_1,\dots,g_i$ is
solvable.  This holds if $(k+1)-j \leq (i-3)/2$, since then the longest
checkerboard is short enough.  If $(k+1)-j \geq (i+1)/2$, then the instance is
not solvable.  The only case that requires a little bit of extra work is
$(k+1)-j=(i-1)/2$, since we then must verify whether the longest checkerboard
is at the beginning or the end.  This, however, is easy.  If the longest
checkerboard has length $(i-1)/2$ and is at the beginning or the end, then the
median group of the instance $g_1,\dots,g_i$, i.e., $g_{(i-1)/2}$ must be a
nonsingleton.  If the longest checkerboard is not at the beginning or the end,
then the median group is a singleton.  This can be tested in constant time.
Hence we can test in amortized constant time whether the instance
$g_1,\dots,g_i$ is solvable.

\begin{corollary}
\label{twocolor_even}
One-column two-color Clickomania with $n$ groups
can be decided in time $O(n)$ if $n$ is even.
If the problem is solvable,
a solution can also be found in time $O(n)$. 
\end{corollary}

\section{Hardness for 5 Colors and 2 Columns}
\label{Hardness for 5 Colors and 2 Columns}

\begin{theorem}
Deciding whether a Clickomania puzzle can be solved
is NP-complete, even if we have only two columns and five colors. 
\end{theorem}

It is relatively easy to reduce two-column six-color Clickomania from
the weakly NP-hard set-partition problem: given a set of integers, can it be
partitioned into two subsets with equal sum?  Unfortunately this does not prove
NP-hardness of Clickomania, because the reduction would represent the integers
in unary (as a collection of blocks).  But the partition problem is only
NP-hard for integers that are superpolynomial in size, so this reduction would
not have polynomial size.  (Set partition is solvable in pseudo-polynomial time,
i.e., time polynomial in the sum of the integers \cite{Garey-Johnson-1979}.)

Thus we reduce from the 3-partition problem, which is strongly NP-hard
\cite{Garey-Johnson-1975, Garey-Johnson-1979}.

\vspace*{-1ex}
\paragraph{3-Partition Problem.}
Given a multiset $A = \{a_1, \dots, a_n\}$ of $n=3m$ positive integers bounded
by a fixed polynomial in $n$,
with the property that $\sum_{i=1}^n a_i = t m$,
is there a partition of $A$ into subsets $S_1, \dots, S_m$ such that $\sum_{a
\in S_i} a = t$ for all $i$?
\medskip

Such a partition is called a \emph{3-partition}.
The problem is NP-hard in
the case that $t/3 \leq a_i \leq 2t/3$ for all $i$.
This implies that a 3-partition satisfies $|S_i| = 3$ for all $i$,
which explains the name.

\begin{figure}[htbp]
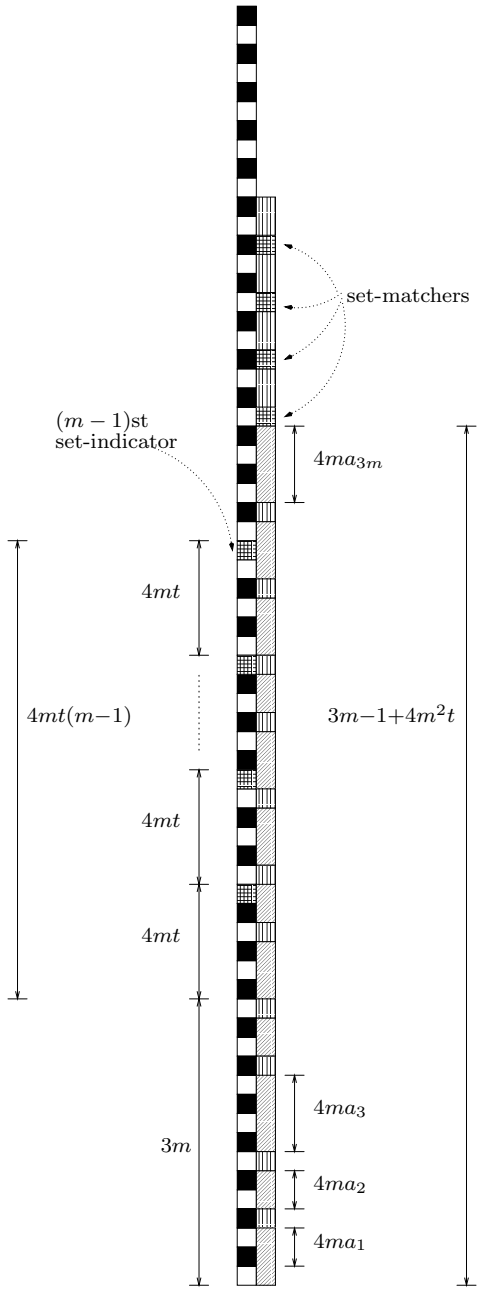

  \centerline{\input overall.pstex_t}
  \caption{\label{overall} Overall construction, not to scale.}
  \end{figure}

The construction has two columns; refer to Figure \ref{overall}.
The left column encodes the sets $S_1,\dots,S_m$ (or more
precisely, the sets $U_j=S_1\cup\dots\cup S_j$  for $j=1,\dots,m-1$,
which is equivalent).  The right column encodes
the elements $a_1,\dots,a_{3m}$, as well as containing separators and
blocks to match the sets.

Essentially, the idea is that in order to remove the singleton
that encodes set $U_j$, we must remove three blocks that encode
elements in $A$, and these elements exactly sum to $t$, hence
form the set $S_j$.

The precise construction is as follows.  The left column consists, from
bottom to top, of the following:
\begin{itemize}
\item $3m$ squares, alternately black and white
\item $m-1$ sections for the $m-1$ sets $U_1,\dots,U_{m-1}$,
	numbered from bottom to top.
	The section for $U_j$ consists of ${\nolinebreak 4mt-1}$ black
	and white squares, follows by one ``red'' square
	(indicated hashed in Figure~\ref{overall}).
	This red square is called the {\em $j$th set-indicator}.
	
	The black and white squares are colored 
	alternatingly black and white, even across
	a set-indicator.  That is, if the last square below a
	set-indicator is white, then the first one above it is
	black and vice versa.
\item Another long stretch of alternating black and white squares.
	There are exactly as many black and white squares above
	the last set-indicator as there were below, and they are
	arranged in such a way that if we removed all set-indicators,
	the whole left column could collapse to nothing.
\end{itemize}

The right column contains at the bottom the elements in $A$, and at
the top squares to remove the set-indicators.  More precisely, the
right column consists, from bottom to top, of the following:
\begin{itemize}
\item $3m$ sections for each element in $A$.  The section for $a_i$
	consists of 1 ``blue'' square (indicated with vertical
	lines in Figure~\ref{overall}) and $4ma_i$ ``green'' squares 
	(indicated with diagonals in Figure~\ref{overall}).   Element
	$a_1$ does not have a separator.
	
	The blue squares are called {\em separators}, while the
	green squares are the one that encode the actual elements.
\item $m-1$ sections for each set.  These consist of three squares each,
	one red and two blue.  The red squares will also be called
	{\em set-matchers}, while the blue squares will again be
	called {\em separators}.  
\end{itemize}

The total height of the construction is bounded by $8m^2t+6m$, which
is polynomial in the input.
And it is not difficult to see that solutions to the puzzle correspond
uniquely to solutions to the 3-partition problem.

\section{Hardness for 3 Colors and 5 Columns}
\label{Hardness for 3 Colors and 5 Columns}

\begin{theorem}
Deciding whether a Clickomania puzzle can be solved
is NP-complete, even if we have only five columns and three colors. 
\end{theorem}

The proof is by reduction from 3-SAT.
We now give the construction.

Let $F=C_1\wedge\cdots\wedge C_m$ be a formula in conjunctive
normal form with variables $x_1,\ldots,x_n$.
We will construct a 5-column Clickomania puzzle using three colors,
white, gray, and black,
where the two leftmost columns, the {\em v-columns},
represent the variables, and the three rightmost columns,
the {\em c-columns}, represent the clauses (see Fig.~\ref{fig_all}(a)).
Most of the board is white, and gray blocks are only used in the
c-columns.
In particular, a single gray block sits on top of the
fourth column, and another white block on top of the gray block.
We will show that this gray block can be removed
together with another single gray block in the rightmost column
if and only if there is a satisfying assignment for $F$.

  \begin{figure}[htbp]
  \centerline{\input all.tex}
  \caption{\label{fig_all} The Clickomania puzzle. The white area is not drawn to scale.}
  \end{figure}%

All clauses occupy a rectangle $CB$ of height $h_{CB}$.
Each variable $x_i$ occupies a rectangle $V_i$ of height $h_v$.
The variable groups are slightly larger
than $CB$, namely $h_v=h_{CB}+3h_0$.
The lowest group $V_0$ represents a dummy
variable $x_0$ with no function other than elevating $x_1$
to the height of $CB$.
The total height of the construction is therefore approximately
$(n+1)\cdot (h_v+3h_0)$.

For all $i$, there are two {\em sliding groups}
$S_{i+1}$ and $\bar{S}_{i+1}$ of size $2h_0$ and $h_0$, respectively,
underneath $V_i$;
their function will be explained later.
The variable groups and the sliding groups are separated by
single black rows
which always count for the height of the group below.
The variable groups contain some more black blocks in the second
column to be explained later.

$CB$ sits above a gray rectangle of height $h_v$ at the bottom
of the c-columns, a white row with a black block in the middle,
a white row with a gray block to the right, and a white rectangle
of height $6h_0-2$.
Fig.~\ref{fig_all}(b) shows the board after we have
removed $V_0,\ldots,V_{i-2}$ from the board,
i.e., assigned a value to the first $i-1$ variables.

$CB$ and $V_i$ are divided into $m$ chunks of height $h_c$,
one for each clause (see Fig.~\ref{fig_all}(c)).
Note that $V_i$ is larger than $CB$, so it also has
a completely white rectangle on top of these $m$ chunks. 
Each clause contains three {\em locks}, corresponding to 
its literals, each variable having a different lock
(we distinguish between different locks by their position
within the clause, otherwise the locks are indistinguishable).
Each variable group $V_i$ on the other hand contains matching
{\em $x_i$-keys} which can be used to open a lock, thus satisfying
the clause.
After we have unlocked all clauses containing $x_i$
we can slide $V_i$ down by removing the white area of $V_{i-1}$
which is now near the bottom of the v-columns.
Thus we can satisfy clauses using all variables, one after the other.

Variables can appear as positive or negative literals,
and we must prevent $x_i$-keys from opening both positive
and negative locks.
Either all $x_i$-keys must be used to open only $x_i$-locks
(this corresponds to the assignment $x_i=1$),
or they are used to open only $\bar{x}_i$-locks (this corresponds
to the assignment $x_i=0$).
To achieve this we use the sliding groups
$S_i$ and $\bar{S}_i$.
Initially, a clause containing literal $x_i$ has its
$x_i$-lock $2h_0$ rows below the $x_i$-key;
if it contains the literal $\bar{x_i}$ then
the $x_i$-lock is $h_0$ rows below the $x_i$-key;
and if it does not contain the variable $x_i$
there is no $x_i$-lock.
So before we can use any $x_i$-key we must slide down $V_i$
by either $h_0$ (by removing $\bar{S}_i$) or by
$2h_0$ (by removing $S_i$).
Removing both $S_i$ and $\bar{S}_i$ slides $V_i$ down by $3h_0$
which again makes the keys useless, so either $x_i=0$
in all clauses or $x_i=1$.

To prevent removal of the large gray rectangle at the bottom
of the c-columns prematurely, we divide each clause into
seven chunks $E_1,\ldots,E_7$ of height $h_0$
each and a barrier group $E_b$ (see Fig.~\ref{fig_all}(d)).
The locks for positive literals are located in $E_5$, and the locks
for negative literals are in $E_6$.
The keys are located in $E_7$.
As said before, we can slide them down by either $h_0$
(i.e., $x=0$), or by $2h_0$ (i.e., $x=1$).
The empty chunks $E_1,\ldots,E_5$ are needed to prevent
misuse of keys by sliding them down more than $2h_0$.

We only describe $E_5$, the construction of $E_6$ is similar
(see Fig.~\ref{fig_all}(e)).
To keep the drawings simple we assume that the v-columns
have been slid down by $2h_0$, i.e., the chunk $E_7$ in the
v-columns is now chunk $E_5$.
$E_5$ is divided into $n$ rectangle $F_1,\ldots,F_n$ of height
$h_1$, one for each variable.
In $V_i$, only $F_i$ contains an
$x_i$-key which is a black rectangle of height
$h_k$ in the second column (see Fig.~\ref{fig_all}(f)), surrounded
on both sides by white space of height $h_s$.
In the c-columns, rectangle $F_{\ell}$ contains an $x_{\ell}$-lock
if and only if the literal $x_{\ell}$ appears in the clause.
The lock is an alternating sequence of black and
white blocks, where the topmost black block is aligned with
the topmost black block of the $x_{\ell}$-key
(see Fig.~\ref{fig_all}(f) and (g)).
The number of black blocks in a lock
varies between clauses, we denote it by $b_j$ for clause $C_j$,
and is independent of the variable $x_i$.
Let $B_j=b_1+\cdots+b_j$.

\begin{figure}[htbp]
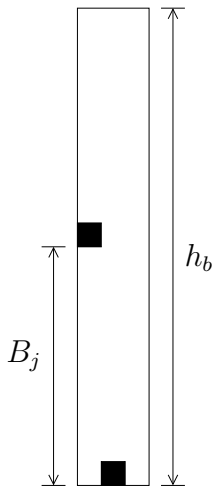

  \centerline{\input barrier.tex}
  \caption{\label{fig_barrier} A barrier in $E_b$}
  \end{figure}

The barrier of clause $C_j$ is located in the chunk $E_b$
of that clause (see Fig.~\ref{fig_barrier}).
It is a single black block in column 4.
There is another single black block in column 3, the {\em bomb},
$B_j$ rows above the barrier.
The rest of $E_b$ is white.
As long as the large white area exists, the only way to remove
a barrier is to slide down a bomb to the same height as the barrier.

With some effort one can show that this board can be solved
if and only if the given formula has a satisfying assignment.

\newpage

\section{Conclusion}
\label{Conclusion}

One intriguing direction for further research is \emph{two-player Clickomania},
a combinatorial game suggested to us by Richard Nowakowski.  In the impartial
version of the game, the initial position is an arbitrary Clickomania puzzle,
and the players take turns clicking on groups with at least two blocks; the
last player to move wins.  In the partizan version of the game, the initial
position is a two-color Clickomania puzzle, and each player is assigned a
color.  Players take turns clicking on groups of their color with at least two
blocks, and the last player to move wins.

Several interesting questions arise from these games.  For example, what is the
complexity of determining the game-theoretic value of an initial position?
What is the complexity of the simpler problem of determining the outcome
(winner) of a given game?  These games are likely harder than the corresponding
puzzles (i.e., at least NP-hard), although they are more closely tied to how
many \emph{moves} can be made in a given puzzle, instead of how many
\emph{blocks} can be removed as we have analyzed here.
The games are obviously in PSPACE, and it would seem natural that they are
PSPACE-complete.

Probably the more interesting direction to pursue is tractability of special
cases.  For example, this paper has shown polynomial solvability of one-column
Clickomonia puzzles, both for the decision and optimization problems.  Can this
be extended to one-column games?  Can both the outcome and the game-theoretic
value of the game be computed in polynomial time?  Even these problems seem to
have an intricate structure, although we conjecture the answers are yes.

In addition, several open problems remain about one-player Clickomania:

  \begin{enumerate}
  \item What is the complexity of Clickomania with two colors?
  \item What is the complexity of Clickomania with two rows?  $O(1)$ rows?
  \item What is the precise complexity of Clickomania with one column?
        Can any context-free-grammar parsing problem be converted into an
        equivalent Clickomania puzzle?  Alternatively, can we construct an
        LR($k$) grammar?
  \item In some implementations, there is a scored version of the puzzle in
        which removing a group of size $n$ results in $(n-2)^2$ points, and
        the goal is to maximize score.  What is the complexity of this problem?
        (This ignores that there is usually a large bonus for removing all
         blocks, which as we have shown is NP-complete to decide.)
  \end{enumerate}

\section*{Acknowledgment}

This work was initiated during the University of Waterloo algorithmic
open problem session held on June 19, 2000.  We thank the attendees of
that meeting for helpful discussions: Jonathan Buss, Eowyn \v Cenek,
Yashar Ganjali, and Paul Nijjar (in addition to the authors).

\bibliography{combinatorialgames,complexity,formallangs}
\bibliographystyle{plain}

\end{document}